          [1999/12/02 1.01 (PWD)] 
\documentclass[a4paper,twocolumn]{esapub} 
\usepackage{natbib} 
 
\title{PROPAGATION OF GAMMA-RAYS IN MASSIVE BINARY CEN X-3: 
INTERACTION OF CASCADE GAMMA-RAYS WITH THE MASSIVE STAR}  
\author{W. Bednarek} 
\affil{Department of Experimental Physics, University of \L
\'od\'z, ul. Pomorska 149/153, 90-236 \L \'od\'z, Poland} 
\affil{bednar@krysia.uni.lodz.pl}

\begin{document} 
 
\keywords{gamma-rays; propagation; massive binaries: Cen X-3}   
\maketitle 
 
\begin{abstract} 
We consider the propagation of very high energy (VHE) gamma-rays in the
radiation field of the massive star of binary system Cen X-3.
VHE gamma-rays or electrons, injected by the compact object close to the
surface of a massive  companion, develop inverse Compton $e^\pm$ 
pair (ICS) cascades. Based on the
Monte Carlo simulations, we obtain the fraction of secondary cascade 
photons which collide with the surface of the massive companion in 
Cen X-3 system. The distribution of photons falling on the surface of
the massive star is investigated. These photons interact
with the atmosphere of the star and should excite different $\gamma$-ray
lines (nuclear, $e^\pm$ annihilation  line).
We estimate that these $\gamma$-ray lines can be likely detected by the
INTEGRAL telescopes. 
\end{abstract} 
 
\section{Introduction} 

Observations of some massive binaries suggest that they may be 
sources of $\gamma$-rays in the GeV and TeV energy range.
For example the massive X-ray binary system, Cen X-3, containing a neutron
star with a 4.8 s period in a 2.09 day orbit around an O-type 
supergiant, has been detected above $100$ MeV $\gamma$-rays by the 
EGRET detector on the Compton Observatory (Vestrand et al.~1997).
There are evidences that observed $\gamma$-ray emission at these 
energies has a form of outbursts and is modulated with a 4.8 s period 
of the pulsar. Based on  the observations in late 80's,
two groups (Brazier et al.~1990, North et al.~1990) reported
Cherenkov detection of positive signal at TeV energies from Cen X-3 at an
orbital  phase of $\sim 0.75$, which is modulated with a period of the pulsar. 
This emission has been localized by Raubenheimer 
\& Smit~(1997) to a relatively small region between the pulsar orbit 
and the surface of a massive companion which may be the accretion 
wake or the limb of the star. More recently the Durham group has 
detected
a persistent flux of $\gamma$-rays above 400 GeV on a lower level than
previous reports (Chadwick et al.~1998,1999a,b). No evidence of correlation 
with the pulsar or orbital periods has been found and no evidence of 
correlation with the X-ray flux has been detected (Chadwick et 
al.~1999a,b).

The problem arises if such TeV photons injected close to the
surface of the massive star (e.g. by the compact object) can 
escape from the soft radiation field of the massive star in the 
binary system Cen X-3. The computations of the optical
depth for TeV photons in the radiation fields of such type massive 
stars show that the absorption of TeV photons should be
significant (e.g. Protheroe \& Stanev~1987, Moskalenko et al.~1993, 
Bednarek
1997).   Recently we have performed Monte Carlo simulations of cascades
initiated by $\gamma$-ray photons and electrons with different energy and
angular distributions, which are injected by the discrete source 
in the radiation field  characteristic
for the Cen X-3 system (Bednarek~2000). The spectra of  escaping $\gamma$-ray
photons and their light curves have been obtained  and discussed in the
context of recent GeV and TeV observations. In this paper we compute 
the $\gamma$-ray power and spectra of photons which fall onto the 
surface of the massive star in Cen X-3.  These photons
should excite different $\gamma$-ray nuclear lines and also 
$e^\pm$ annihilation line. Their detection by INTEGRAL telescopes
should give independent information on high energy processes which
seems to occur in Cen X-3 system.   
The following parameters for the massive binary Cen X-3 are assumed: 
the radius of the O star $r_{\rm s} = 8.6\times 10^{11}$ cm, its 
surface temperature $T_{\rm s} = 3\times 10^4$ K, the binary star
separation $a = 1.2\times 10^{12}$ cm (Krzemi\'nski~1974).

\begin{figure} 
  \vspace{6.cm} 
  \includegraphics{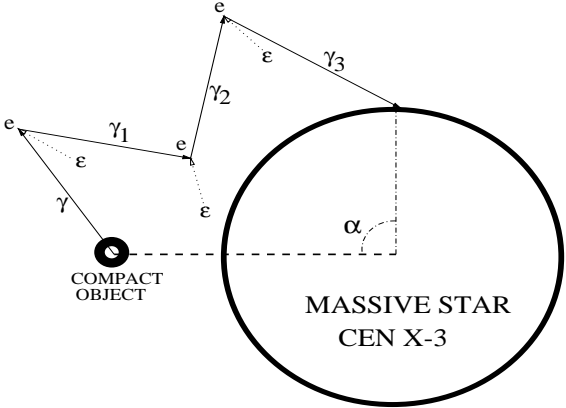}
  \caption[]{Schematic picture of the cascade initiated by a 
primary gamma-ray
($\gamma$) in the radiation field of a massive star. Gamma-ray photon,
injected by the compact object or produced by primary electron, 
creates $e^\pm$ pairs (e) in the interaction with a soft star photon
$\varepsilon$. The secondary $e^\pm$ pairs, by scattering soft star
photons, create secondary gamma-rays ($\gamma_1$, $\gamma_2$, 
$\gamma_3$). Some of them ($\gamma_3$) may fall onto the surface of the 
massive star.}
\label{Fig1}
\end{figure} 

%
%
\section{Gamma-rays from ICS cascade} 

Let us assume that $\gamma$-ray photons are injected by the compact
object which is on an orbit around the star with the parameters 
characteristic for Cen X-3 system. These photons may 
create $e^\pm$ pair in collision with the soft star photon 
(Bednarek~2000).
The secondary pairs can next produce ICS $\gamma$-rays, initiating in 
this way the ICS cascade in the massive star radiation which 
is seen anisotropic in
respect to the location of the injection place of the primary photons
or electrons. 

\begin{figure*}[t] 
  \vspace{6.5cm} 
 \includegraphics{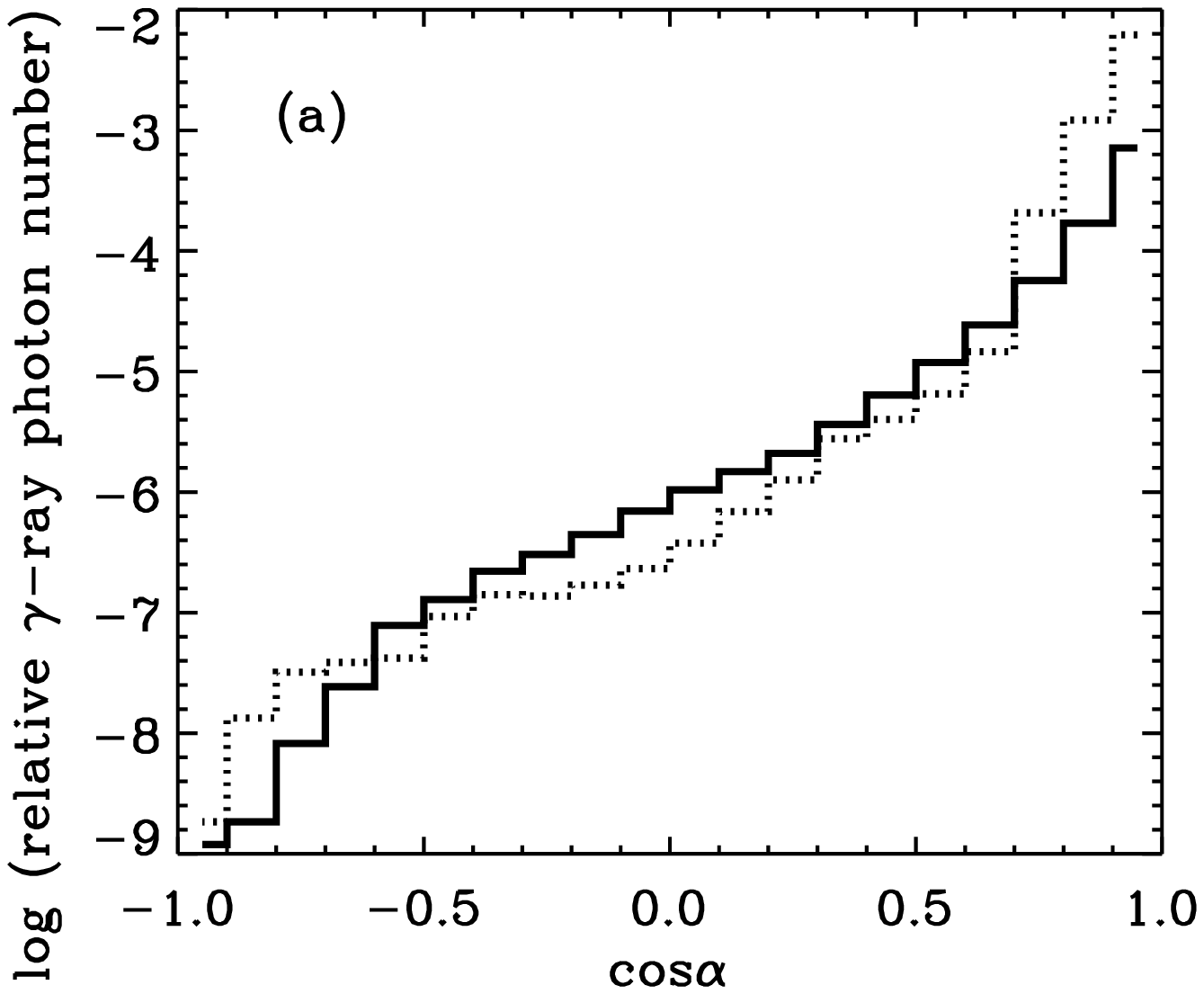}
 \includegraphics{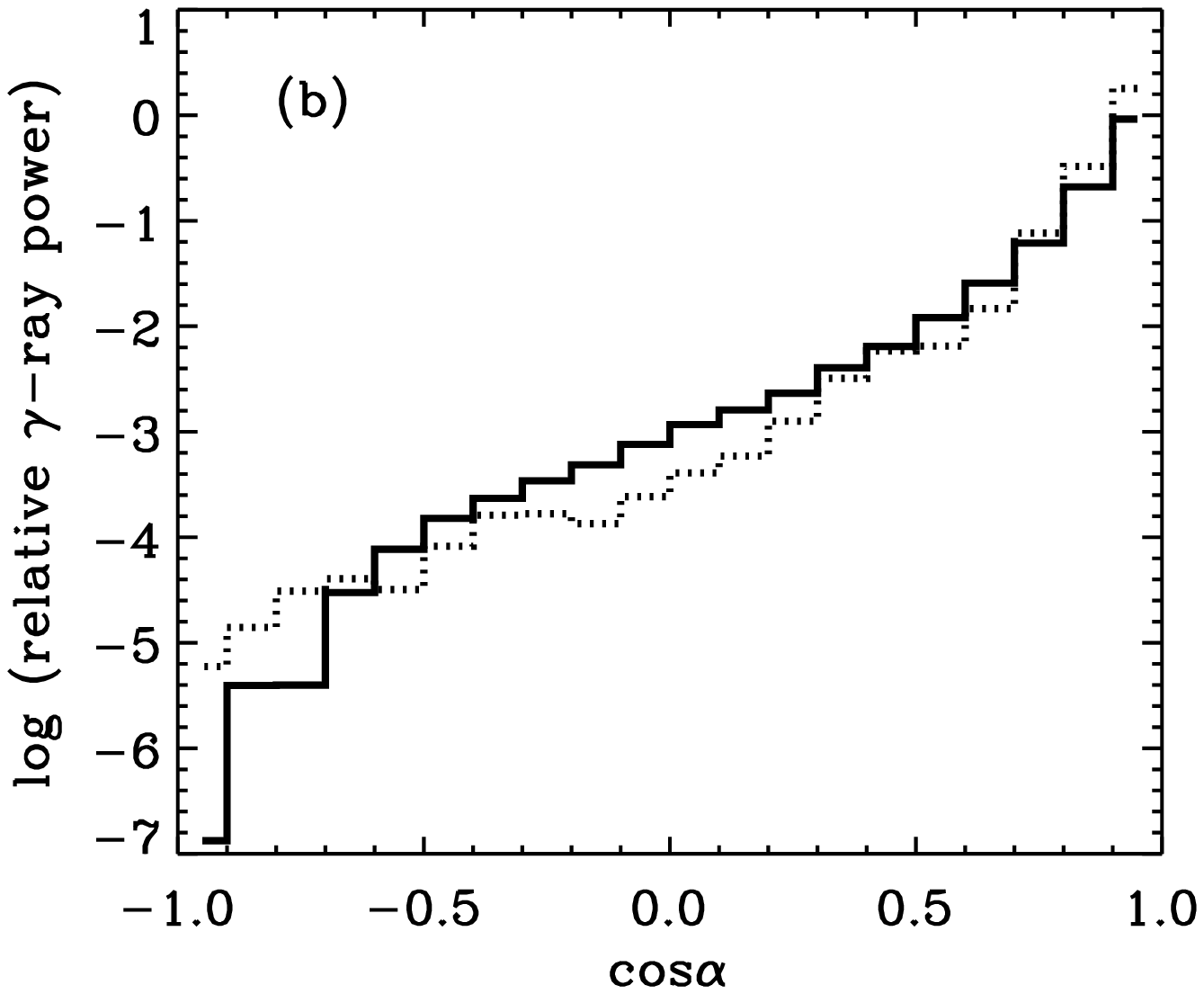}
\caption[]{The number of $\gamma$-rays {\bf (a)} and $\gamma$-ray power
{\bf (b)} falling  on the surface of the massive star in Cen X-3 system. 
It is assumed that the point like source injects isotropically primary
$\gamma$-rays (dotted histogram) or primary electrons (full) with a 
power law spectra and spectral index -2.}  
\label{Fig2}  
\end{figure*} 

The propagation of secondary pairs inside the binary 
system is determined by the strength and the geometry of the magnetic 
field 
which in general may be very complicated. We analize the case in which 
the random component of the magnetic field isotropize secondary cascade 
$e^\pm$
pairs close to the place  of their creation before they next scatter soft
photon from the massive star. This condition is met if the mean free path for
ICS process is  longer than the Larmor radius of secondary pairs in the local
magnetic  field (see discussion and condition for that to occur in
Bednarek~1997).  We neglect also the synchrotron losses of secondary  pairs in
respect  to their ICS losses on star photons. This can be done  for the values 
of the magnetic field below certain limit (see Eq.~3 in Bednarek~1997). 
The bremsstrahlung losses of pairs in matter with the star wind 
densities characteristic for the OB stars are negligible.

In the previous paper (Bednarek~2000), we show the details of the
Monte Carlo simulations and concentrate on the photons which escape
from the binary system. However significant part of secondary cascade 
photons should collide with the massive star. The distribution of these 
photons on the star surface, their spectra and the fraction of photon
power which fall onto the surface in respect to the power of all 
produced cascade photons is investigated in this paper. The considered
picture is shown in Fig.~\ref{Fig1}. 

We concentrate on two cases which were already discussed in 
Bednarek~(2000), i.e. the case of isotropic injection of electrons and
isotropic injection of photons with a power law spectrum and
index -2. Electrons with such distribution can be accelerated at the 
shock front created in collision of the pulsar wind with the surrounding
matter as proposed in the model by Kennel \& Coroniti~(1984). If the 
compact object (neutron star) is surrounded by dense cocoon created 
by matter accreting spherically-symmetric or by the
accretion disk, then the $\gamma$-ray photons should be injected into 
the volume of the binary system. Based on our simulations we show 
in Figs.~\ref{Fig2}{\bf a} and {\bf b}
the distribution of secondary photons on the surface of the massive star 
and the distribution of the $\gamma$-ray power of these secondary 
photons as a function of the cosine angle $\alpha$, measured from the
direction  defined by the centers of the massive star and the compact 
object. Note that this distribution is axially symmetric (see Fig.~1).

\begin{figure*} 
  \vspace{6.5cm} 
 \includegraphics{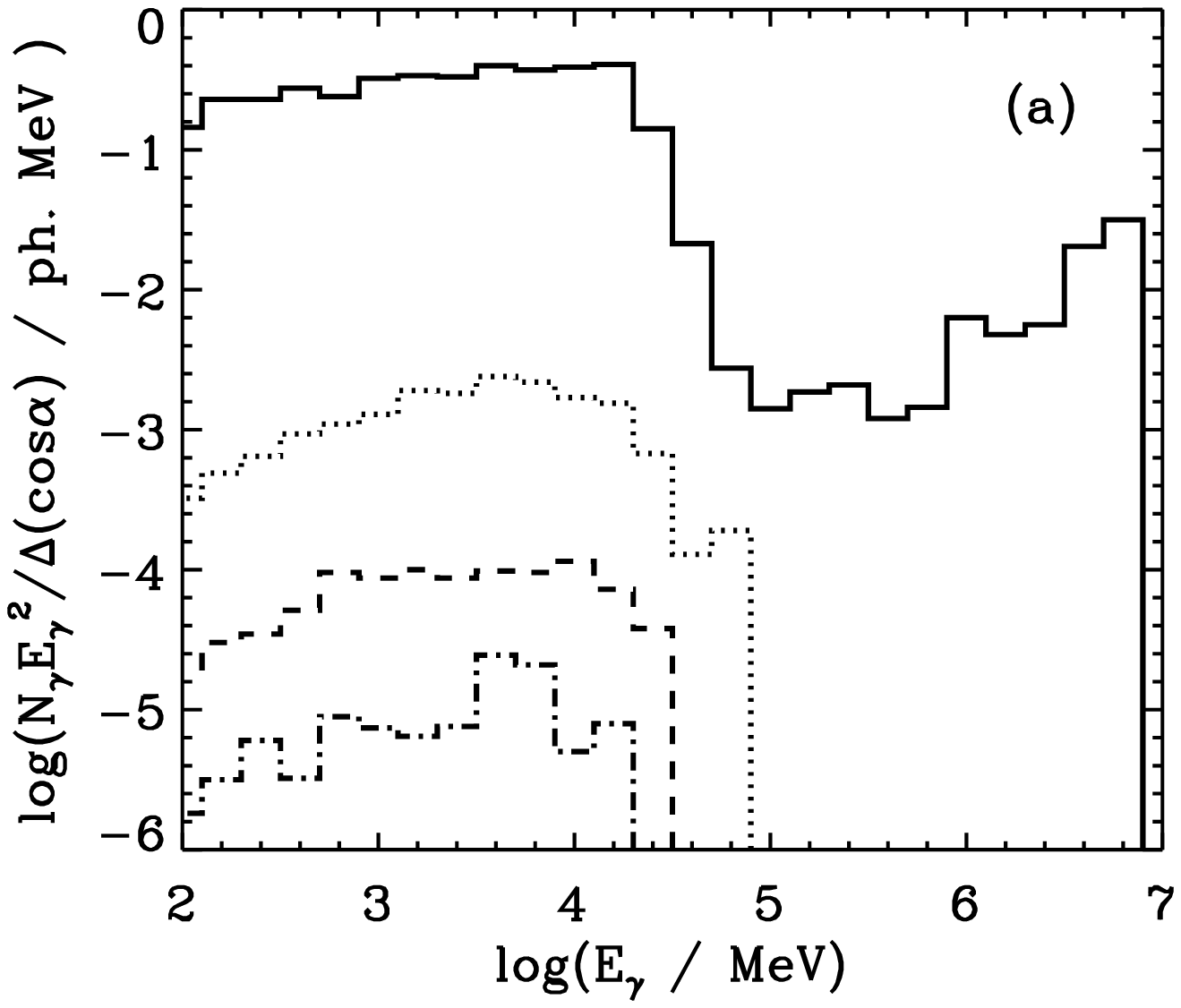}
 \includegraphics{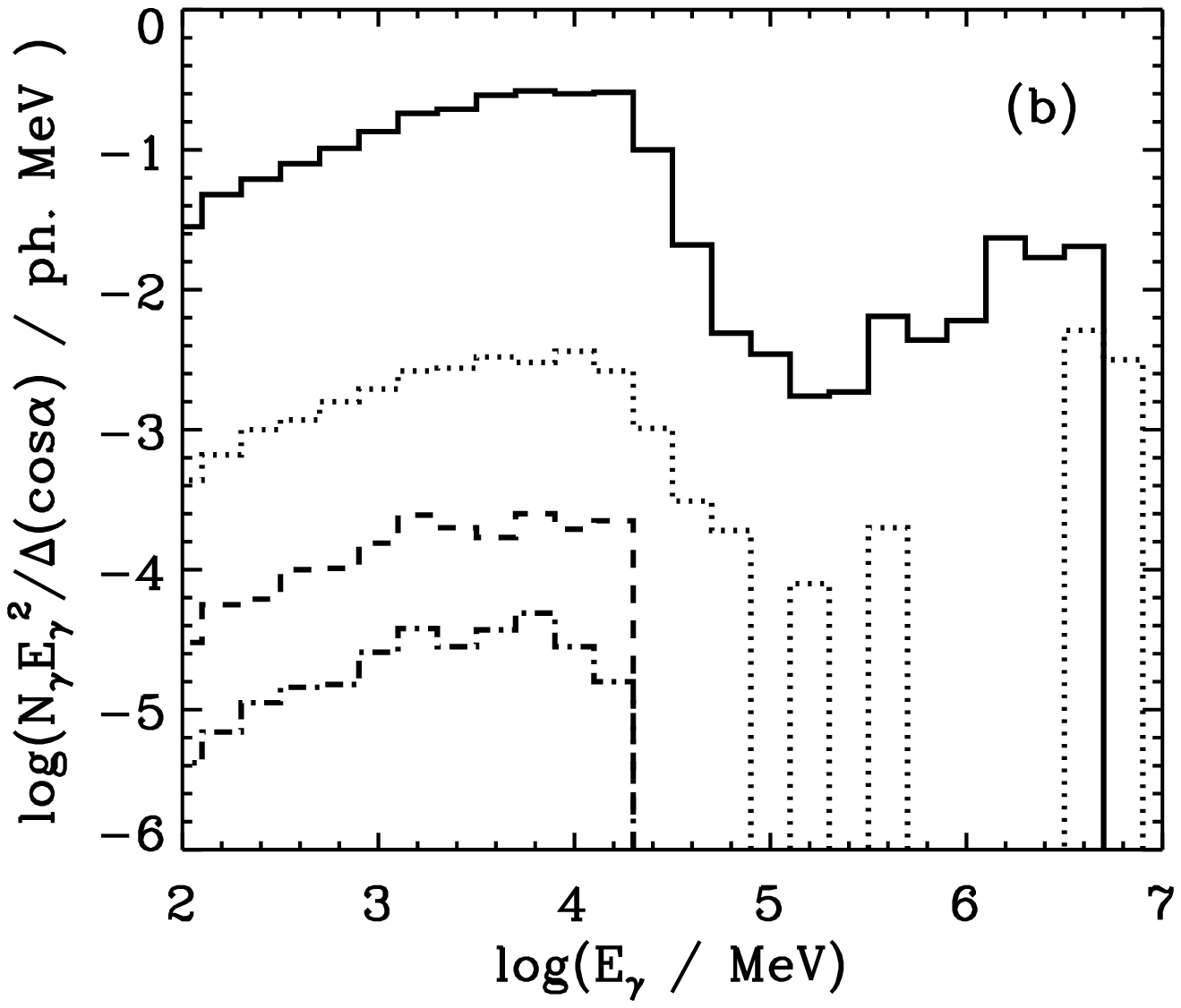}
\caption[]{The spectra of $\gamma$-ray photons falling onto the
surface of the massive star in Cen X-3 system within the cosine angle
$\alpha$: 
$\cos\alpha = 0.9\leftrightarrow 1$ (full histogram), 
$0.5\leftrightarrow 0.6$ (dotted),
$0.\leftrightarrow 0.1$ (dashed), and 
$-0.5\leftrightarrow -0.4$ (dot-dashed). The primary
electrons {\bf (a)} and photons {\bf (b)}, with a power law  spectrum 
and spectral index -2, are injected isotropically by the compact 
object at the distance of 1.4 radii from the massive star. }      
\label{Fig3}  
\end{figure*} 

The numbers of secondary photons and their power 
falling onto the massive star drop strongly with the angle $\alpha$.
Only a small number of photons is able to fall onto the opposite 
side of the massive star than the location of the compact object 
(the source of primary particles initiating the cascade), 
i.e. at cosine angles less than $\cos\alpha = 0$. This angular 
distribution of falling photons is much sharper than the angular 
distribution of cascade photons escaping from the system. 
These interesting features are due to the  type of the considered
cascade process in which the secondary $e^\pm$ pairs are isotropized by 
the random magnetic field. These pairs produce secondary ICS photons 
moving towards the massive star surface with higher probability. 
However for secondary photons it is easier to escape from the system 
on the oposite side of the massive star than collide with the surface
of the massive star on the opposite side.
There are also
differences in distributions of secondary photons on the massive 
star surface in the case of injection
of primary photons and electrons (compare dotted and full histograms in    
Figs.~\ref{Fig2}{\bf a} and {\bf b}). The distribution of secondary 
photons from cascades initiated by primary photons shows higher number 
of secondary photons falling onto the opposite side of the massive star 
and on the surface of massive star just below the compact object the 
than in the case of injection of primary electrons.  This is due to the 
fact that primary photons produce first generation of $e^\pm$ pairs in 
quite extended region around the compact object in comparison to the 
point like injection of primary electrons in the second case. We show 
also in Figs.~\ref{Fig3}{\bf a} and {\bf b}  the spectra of secondary 
photons falling onto the massive star in both discussed cases for 
different range of cosine angles $\alpha$.  As expected, the secondary
photons with the highest energies fall mainly onto the massive star for 
small angles $\alpha$. On the opposite side of the massive star fall 
only secondary photons with relatively low  energies. 

We estimate also the ratio of power of secondary photons which 
fall onto the massive star to power of photons which escape from the 
system.
In the case of injection of primary photons this ratio is equal to
$\sim 13\%$  and in the case of injection of primary electrons it is 
equal to $\sim 40\%$. In the case of primary electrons, this is much 
more than expected from simple geometrical considerations (the case of 
non-cascading photons) which
predicts that only $\sim 13\%$ of photons should fall into the star 
surface since this part of the sky is shaded by the massive star.
In the case of primary photons, the non-cascading photons dominate
and the ratios are comparable. 

This results show that significant part of the $\gamma$-ray power
observed from the binary system should be transferred onto the surface 
of the massive star. These $\gamma$-rays can heat the star and partially 
can be re-emitted in the form of nuclear lines and $e^\pm$ annihilation 
line. Therefore we suggest that the binary systems of the Cen X-3 type 
can also produce $\gamma$-ray lines in the MeV energy range. 

\begin{figure*} 
  \vspace{6.5cm} 
 \includegraphics{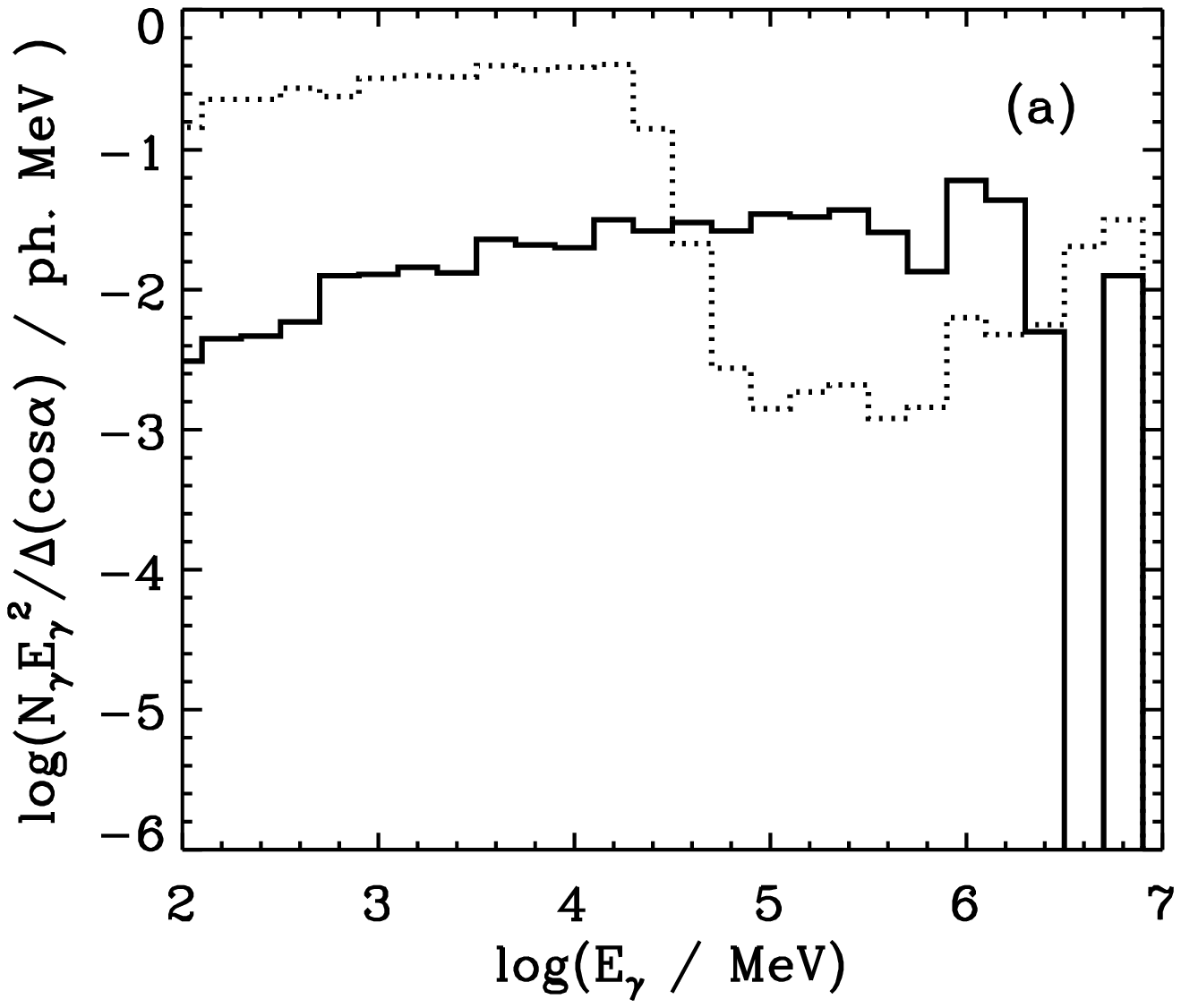}
 \includegraphics{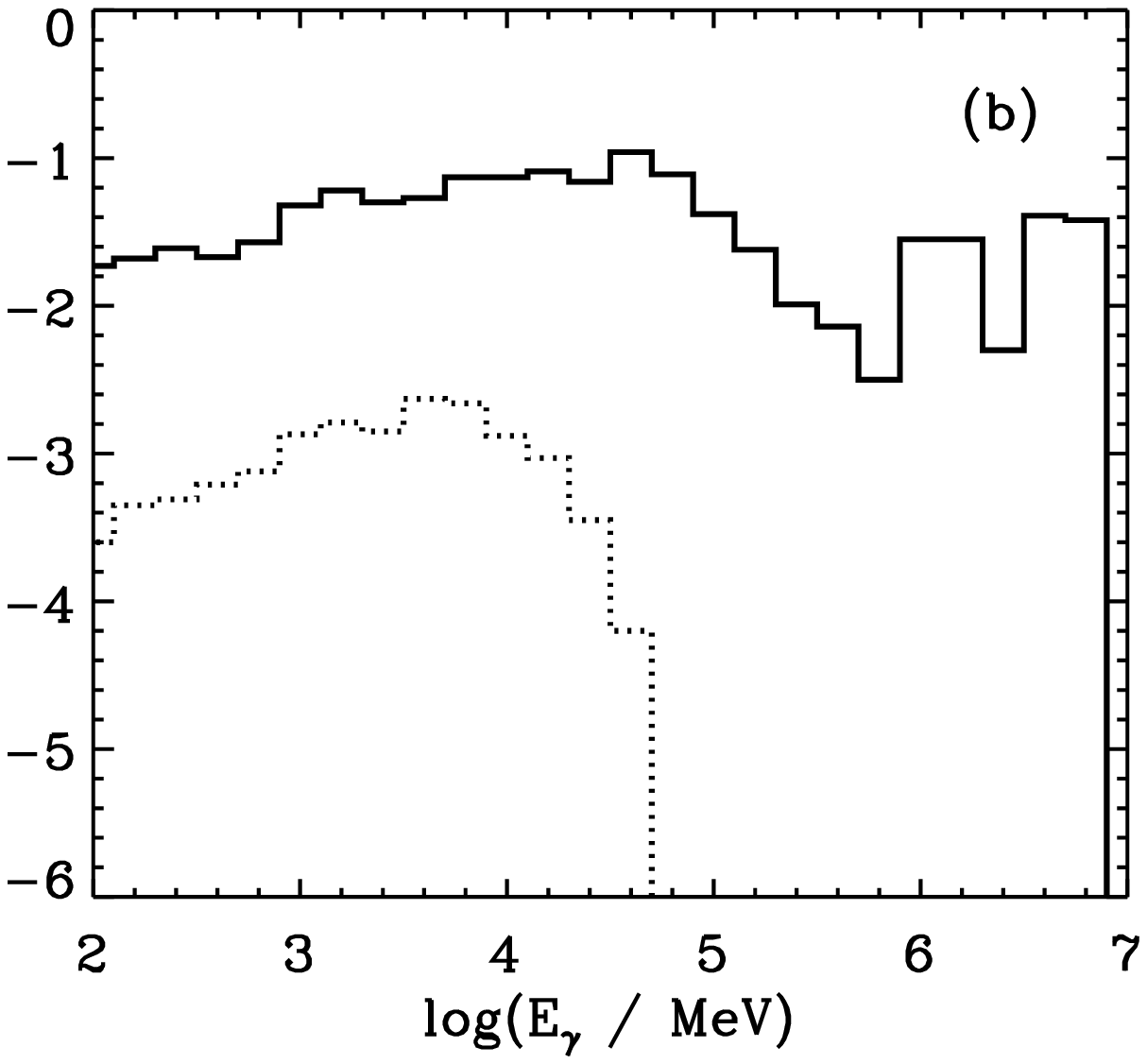}
 \includegraphics{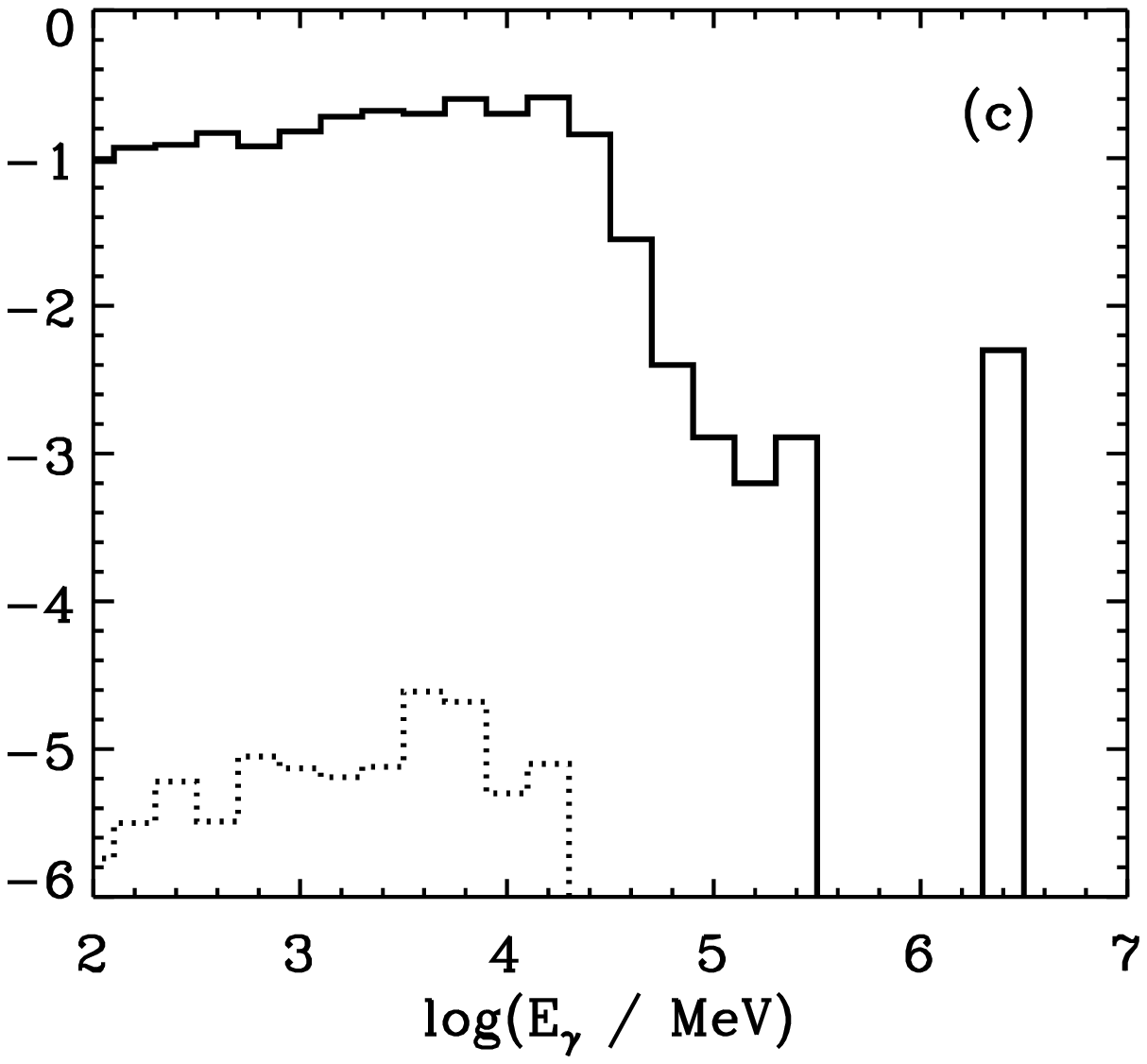}
\caption[]{Comparison of the secondary $\gamma$-ray spectra which escape
from the binary system Cen X-3 (full histograms) and which fall onto the 
surface of the massive star (dotted histograms) for these same cosine
angles $\alpha$ equal to $\cos\alpha  = 0.9\leftrightarrow 1.$ {\bf (a)}, 
$0.5\leftrightarrow 0.6$ {\bf (b)}, and $-0.5\leftrightarrow -0.4$ {\bf (c)}.
The  primary electrons with a power law spectrum and spectral index -2. 
are injected by the compact object.}  
\label{Fig4}  
\end{figure*}

To find out what is the relation between this
expected $\gamma$-ray line emission and the continuous emission, we
compare in Figs.~\ref{Fig4}{\bf a,b} and {\bf c} the spectra of photons 
which fall onto the surface of the massive star at fixed angles $\alpha$ 
with the spectra of secondary cascade photons which escape from the
system at these same ranges of angles
(see Bednarek~2000). Fig.~\ref{Fig4}{\bf a} shows that the line 
emission should be the strongest when the compact object is in front of 
the massive star. Similar behaviour is expected for the escaping cascade 
$\gamma$-rays with energies above 100 GeV (ground Cherenkov telescopes).
On the contrary, the highest fluxes of 
escaping cascade photons with energies above 100 MeV (space telescopes) 
are expected in the direction tangent to the massive star limb
(see Fig.~\ref{Fig4}{\bf c}).
Therefore it is expected that the maximum of the $\gamma$-ray line 
light curve should not correspond to the maximum of the 100 MeV 
$\gamma$-ray continuous emission.

%
%
\section{Conclusion} 

Assuming that VHE $\gamma$-ray photons or electrons are injected 
inside the binary system Cen X-3, as postulated by observations
of this source in the GeV and TeV energy ranges, we consider the 
cascades initiated by these particles in the radiation field of the 
massive star in Cen X-3. The spectra of secondary photons, which escape 
from the system and fall onto the surface of the massive star, are 
obtained by using the Monte Carlo method. The photons which fall onto
the massive star should excite the $\gamma$-ray nuclear lines and the
$e^\pm$ annihilation line. The maximum intensities of $\gamma$-ray line
emission is expected when the compact object is in front of the massive 
star. On the contrary, the maximum emission of the 100 MeV emission
is expected for direction tangent to the limb of the massive star.
Therefore the $\gamma$-ray light curves observed at low and high energies
should appear different.

Our model predicts that about $13\leftrightarrow 
40\%$ of $\gamma$-ray power 
escaping
from  the system can be transferred to the surface of the massive star as 
a result of cascades in the binary system Cen X-3. The phase averaged
$\gamma$-ray luminosity above 100 MeV, reported by the EGRET telescope, 
is $\sim 5\times 10^{36}$ erg s$^{-1}$ (Vestrand et al.~1997).
Therefore about $\sim 1\leftrightarrow 
2\times 10^{36}$ erg s$^{-1}$ should
fall  onto the
star. The INTEGRAL Spectrometer instrument will have narrow-line 
sensitivity of 
a $5\times 10^{-6}$ ph cm$^{-2}$ s$^{-1}$ at 1 MeV and 
continuum sensitivity
$1.5\times  10^{-7}$ ph cm$^{-2}$ s$^{-1}$ at 1 MeV 
(http://astro.estec.esa.nl/SA-general/projects/INTEGRAL).
It will be able to detect the 1 MeV $\gamma$-ray source emitting the 
power $\sim 10^{35}$ erg s$^{-1}$ at the distance of Cen X-3 system
equal to 8 kpc (Krzemi\'nski~1974). Therefore the conversion efficiency 
of cascade photons into the $\gamma$-ray line photons on the surface of 
the massive star should be $\sim 3\leftrightarrow 10\%$.
This value does not seems to be extraordinary.
The detailed predictions for possible detection of such lines from 
Cen X-3 by the INTEGRAL telescopes will be considered in the future.

\section*{Acknowledgments} 
This work is supported by the grant from the University of 
\L \'od\'z  No. 505/703.

\end{document}